\documentclass[11pt]{article}
\usepackage{graphicx}

\catcode`\@=11
\long\def\@makefntext#1{
\protect\noindent \hbox to 3.2pt {\hskip-.9pt
$^{{\eightrm\@thefnmark}}$\hfil}#1\hfill}       %CAN BE USED

\def\@makefnmark{\hbox to 0pt{$^{\@thefnmark}$\hss}}    %ORIGINAL

\def\ps@myheadings{\let\@mkboth\@gobbletwo
\def\@oddhead{\hbox{}
\rightmark\hfil\eightrm\thepage}
\def\@oddfoot{}\def\@evenhead{\eightrm\thepage\hfil
\leftmark\hbox{}}\def\@evenfoot{}
\def\sectionmark##1{}\def\subsectionmark##1{}}

\oddsidemargin=\evensidemargin
\newcounter{sectionc}\newcounter{subsectionc}\newcounter{subsubsectionc}
\renewcommand{\section}[1] {\vspace{12pt}\addtocounter{sectionc}{1}
\setcounter{subsectionc}{0}\setcounter{subsubsectionc}{0}\noindent
    {\tenbf\thesectionc. #1}\par\vspace{5pt}}
\renewcommand{\subsection}[1] {\vspace{12pt}\addtocounter{subsectionc}{1}
    \setcounter{subsubsectionc}{0}\noindent
    {\bf\thesectionc.\thesubsectionc. {\kern1pt \bfit #1}}\par\vspace{5pt}}
\renewcommand{\subsubsection}[1] {\vspace{12pt}\addtocounter{subsubsectionc}{1}
    \noindent{\tenrm\thesectionc.\thesubsectionc.\thesubsubsectionc.
    {\kern1pt \tenit #1}}\par\vspace{5pt}}

\def\eightcirc{
\begin{picture}(0,0)
\put(4.4,1.8){\circle{6.5}}
\end{picture}}
\def\eightcopyright{\eightcirc\kern2.7pt\hbox{\eightrm c}}

\def\abstracts#1#2#3{{
    \centering{\begin{minipage}{4.5in}\baselineskip=10pt\footnotesize
    \parindent=0pt #1\par
    \parindent=15pt #2\par
    \parindent=15pt #3
    \end{minipage}}\par}}

\def\keywords#1{{
    \centering{\begin{minipage}{4.5in}\baselineskip=10pt\footnotesize
    {\footnotesize\it Keywords}\/: #1
     \end{minipage}}\par}}

\renewenvironment{thebibliography}[1]
    {\frenchspacing
     \ninerm\baselineskip=11pt
     \begin{list}{\arabic{enumi}.}
        {\usecounter{enumi}\setlength{\parsep}{0pt}
     \setlength{\leftmargin 12.7pt}{\rightmargin 0pt} %FOR 1--9 ITEMS
         \setlength{\itemsep}{0pt} \settowidth
    {\labelwidth}{#1.}\sloppy}}{\end{list}}

\newcounter{itemlistc}
\newcounter{romanlistc}
\newcounter{alphlistc}
\newcounter{arabiclistc}

\def\@citex[#1]#2{\if@filesw\immediate\write\@auxout
    {\string\citation{#2}}\fi
\def\@citea{}\@cite{\@for\@citeb:=#2\do
    {\@citea\def\@citea{,}\@ifundefined
    {b@\@citeb}{{\bf ?}\@warning
    {Citation `\@citeb' on page \thepage \space undefined}}
    {\csname b@\@citeb\endcsname}}}{#1}}

\newif\if@cghi
\def\cite{\@cghitrue\@ifnextchar [{\@tempswatrue
    \@citex}{\@tempswafalse\@citex[]}}
\def\citelow{\@cghifalse\@ifnextchar [{\@tempswatrue
    \@citex}{\@tempswafalse\@citex[]}}
\def\@cite#1#2{{$\null^{#1}$\if@tempswa\typeout
    {IJCGA warning: optional citation argument
    ignored: `#2'} \fi}}

\def\@refcitex[#1]#2{\if@filesw\immediate\write\@auxout
    {\string\citation{#2}}\fi
\def\@citea{}\@refcite{\@for\@citeb:=#2\do
    {\@citea\def\@citea{, }\@ifundefined
    {b@\@citeb}{{\bf ?}\@warning
    {Citation `\@citeb' on page \thepage \space undefined}}
    \hbox{\csname b@\@citeb\endcsname}}}{#1}}

\def\@refcite#1#2{{#1\if@tempswa\typeout
        {IJCGA warning: optional citation argument
    ignored: `#2'} \fi}}

\def\refcite{\@ifnextchar[{\@tempswatrue
    \@refcitex}{\@tempswafalse\@refcitex[]}}

\def\pmb#1{\setbox0=\hbox{#1}
    \kern-.025em\copy0\kern-\wd0
    \kern.05em\copy0\kern-\wd0
    \kern-.025em\raise.0433em\box0}

\def\fnt#1#2{\footnotetext{\kern-.3em
    {$^{\mbox{\scriptsize #1}}$}{#2}}}

\font\tenrm=cmr10
\font\tenit=cmti10
\font\tenbf=cmbx10
\font\bfit=cmbxti10 at 10pt
\font\ninerm=cmr9

\font\eightrm=cmr8

\def\qed{\hbox{${\vcenter{\vbox{            %HOLLOW SQUARE
   \hrule height 0.4pt\hbox{\vrule width 0.4pt height 6pt
   \kern5pt\vrule width 0.4pt}\hrule height 0.4pt}}}$}}

\begin{document}

\centerline{\bf TALBOT EFFECT FOR DISPERSION IN LINEAR OPTICAL
FIBERS}

\centerline{\bf  AND A WAVELET APPROACH}
%\footnote{
%This essay received an ``honorable mention'' in the
 %       Annual Essay Competition of the Gravity Research
    %    Foundation for the year 2000 - Ed.}}
\vspace*{0.035truein}
%\centerline{\bf MANUSCRIPTS USING COMPUTER SOFTWARE\footnote{For
%the title, try not to use more than 3 lines. Typeset the title
%in 10 pt Times Roman, uppercase and boldface.}}
\vspace*{0.37truein}
%\centerline{\footnotesize NAME}
%\footnote{Typeset names in
%10 pt Times Roman, uppercase. Use the footnote to indicate the
%present or permanent address of the author.}}
%\vspace*{0.015truein}
%\centerline{}
%\baselineskip=10pt
%\centerline{\footnotesize\it City, State ZIP/Zone,
%Country\footnote{State completely without abbreviations, the
%affiliation and mailing address, including country. Typeset in 8
%pt Times Italic.}}
\vspace*{10pt} \centerline{\footnotesize H.C.
ROSU$^{a}$\footnote{E-mail: hcr@ipicyt.edu.mx \hfill {\tt
physics/0510067}}, $\,$
J.P. TREVI\~NO$^{a}$, H. CABRERA$^{a}$,  and J.S. MURGU\'IA$^{b}$}%\footnote{E-mail: ondeleto@hotmail}}
\vspace*{0.015truein}
%\centerline{\footnotesize [Received 17 August 1999] }
%\baselineskip=10pt
\centerline{\footnotesize $a)$ {\em IPICyT} - Instituto Potosino de
Investigaci\'on Cient\'{\i}fica y Tecnol\'ogica,} \centerline{
\footnotesize Apdo Postal 3-74 Tangamanga, 78231 San Luis
Potos\'{\i}, M\'exico} \centerline{ \footnotesize $b)$ Universidad
Aut\'onoma de San Luis Potos\'{\i}, 87545 San Luis Potos\'{\i},
M\'exico} \vspace*{0.225truein}
%%%%%%%%%%%%%%%%%%%%%%%%%%%%%%%%%%%%%%%%%%%%%
%\publisher{(June 2002) - Mod Phys. Lett. A 17}
%{(May 14, 2000)}

%\centerline{\footnotesize Dated Feb 25th 2004 $\qquad$ file: DIRACC.TEX}

%\author{H.C. ROSU, J.P. TREVI\~NO, H. CABRERA}

%\author{J.S. MURGU\'IA}

%\address{Universidad
%Aut\'onoma de San Luis Potos\'{\i}, 87545 San Luis Potos\'{\i},
%M\'exico\\
%ondeleto@uaslp.mx}

%\maketitle

%\begin{history}
%\received{Day Month Year} \revised{Day Month Year}
%\accepted{(Day Month Year)}
%\comby{(xxxxxxxxxx)}
%\end{history}

\abstracts{ We shortly recall the mathematical and physical aspects
of Talbot's self-imaging effect occurring in near-field diffraction.
In the rational paraxial approximation, the Talbot images are formed
at distances $z=p/q$, where $p$ and $q$ are coprimes, and are
superpositions of $q$ equally spaced images of the original binary
transmission (Ronchi) grating. This interpretation offers the
possibility to express the Talbot effect through Gauss sums. Here,
we pay attention to the Talbot effect in the case of dispersion in
optical fibers presenting our considerations based on the close
relationships of the mathematical  representations of diffraction
and dispersion. Although dispersion deals with continuous functions,
such as gaussian and supergaussian pulses, whereas in diffraction
one frequently deals with discontinuous functions, the mathematical
correspondence enables one to characterize the Talbot effect in the
two cases with minor differences. In addition, we apply, for the
first time to our knowledge, the wavelet transform to the fractal
Talbot effect in both diffraction and fiber dispersion. In the first
case, the self similar character of the transverse paraxial field at
irrational multiples of the Talbot distance is confirmed, whereas in
the second case it is shown that the field is not self similar for
supergaussian pulses. Finally, a high-precision measurement of
irrational distances employing the fractal index determined with the
wavelet transform is pointed out.
%\end{abstract}
}{}{}

\bigskip

%\section{General Appearance}    %) A SECTION HEADING
\section{Introduction}

Near field diffraction can produce images of periodic structures
such as gratings without any other means. This is known since 1836
when this self-imaging phenomenon has been discovered by H.F.
Talbot, one of the inventor pioneers of photography.\cite{hft} Take
for example a periodic object as simple as a Ronchi grating which is
a set of black lines and equal clear spaces on a plate repeating
with period $a$. In monochromatic light of wavelength $\lambda$ one
can reproduce its image at a ``focal" distance known as the Talbot
distance given by $z_{\rm T}=a^2 \lambda ^{-1}$ as shown in Fig.~1.
Actually, this famous focal distance has been first derived by Lord
Rayleigh in 1881.\cite{Rayleigh81} Moreover, more images show up at
integer multiples of the Talbot distance. It was only in 1989 that
the first and only one review on the Talbot effect has been written
by Patorski.\cite{pator}

In the framework of Helmholtz equation approach to the physical
optics of a $\delta$-comb grating, Berry and Klein\cite{ref1} showed
in 1996 that the Talbot diffraction wavefield has a highly
interesting arithmetic structure related to Gauss sums and other
fundamental relationships in number theory.

In this paper, after briefly reviewing the results of Berry \&
Klein, we show that analogous results can be obtained easily for the
case of dispersion in linear optical fibers. Moreover, we apply for
the first time in the literature the wavelet transform\cite{primer}
to the fractal Talbot problem. The point with the wavelet transform
is that it contains more information with respect to the Fourier
transform, which is behind the Helmholtz equation. Wavelet
transforms have been considered as a very suitable mathematical
microscope for fractals due to their ability to reveal the
construction rules of fractals and to resolve local scaling
properties as noticed before for the case of fractal
aggregates.\cite{arneodo88}

%22222222222222222222222222222222222222222222222222222222222222222222222222222222222222222222222222222222222222222222222222222222222222222222222222

%...................................................................................   SECTION 2        GS BY HELMHOTZ
\section{Talbot effect for Helmholtz scalar diffraction fields}

The diffraction due to Ronchi gratings can be approached
analytically using the Helmholtz equation. Passing to dimensionless
transverse and paraxial variables $\xi=x/a$ and $\zeta = z/a$,
respectively, the scalar wave solution $\Psi (\xi , \zeta)$ of the
Helmholtz equation can be expressed as a convolution in $\xi$ of the
Ronchi unit cell square function
%....................
\begin{equation}\label{unit cell}
g(\xi) =\left\{\begin{array}{c}
1\quad \xi\in [-\frac{1}{4},\frac{1}{4}]\\
0\quad \xi \ni [-\frac{1}{4},\frac{1}{4}]~,
\end{array}\right.
\end{equation}
%.................
and the Dirac comb transmittance, i.e.
%........................
\begin{equation}\label{scalarH}
\Psi(\xi , \zeta) =\int _{-1/2}^{+1/2} g(\xi ^{'})\left(\sum
_{n=-\infty}^{\infty} \exp [i \, 2\pi n(\xi -\xi ^{'})]\exp [i\Theta
_n(\zeta)]\right)d\xi ^{'}~.
\end{equation}
%.....................................
In the previous formulas, the unit cell is the single spatial period
of the grating, which we take centered at the origin and of length
equal to unity and $\Theta _n(\zeta)=2\pi\zeta \frac{a^2}{\lambda
^2} \sqrt{1-\left(\frac{n\lambda}{a}\right)^2}$ is a phase produced
by the diffraction of the Dirac comb `diagonal' rays. The so-called
Fresnel approximation for this phase, i.e., a Taylor expansion up to
the second order for the square root, leads to $\Theta _s(\zeta)
\approx -\pi n^2\zeta$. It can be easily shown now that in the
Fresnel approximation Eq.~\ref{scalarH} can be written as an
infinite sum of phase exponentials in both variables $\xi$ and
$\zeta$
%.......................
\begin{equation}\label{sum_exps}
\Psi_{\rm p}(\xi , \zeta) = \sum _{n=-\infty}^{\infty} g_n \exp [i\,
2\pi n \xi - i\pi n^2\zeta]=\sum _{n=-\infty}^{\infty} g_n\psi _{\rm
p} (\xi ,
\zeta)~, %\Theta _s(\zeta)]~,
\end{equation}
%..............................
where the amplitudes $g_n$ are the Fourier modes of the
transmittance function of the Ronchi grating
%.....................
\begin{equation}\label{Fourier_g}
g_n=\int _{-1/4}^{+1/4} d\xi ^{'}\exp [-i \, 2\pi n\xi ^{'}]~.
\end{equation}
%........................
%On the other hand, $\Theta _s$ is a phase produced by the diffraction of the Dirac comb `diagonal' rays. The so-called Fresnel
%approximation for  $\Theta _s$, i.e., a Taylor expansion up to the second order for a
%square root coming from the `diagonal' feature of the ray, leads to $\Theta _s \approx -\pi \zeta n^2$.
%.................................................................
%Thus, the following form of the paraxial field is obtained
%...................
%\begin{equation}
%\Psi_p(\xi,\zeta)=\sum _{n=-\infty}^{\infty} g_n \exp (i2\pi n \xi
%-i\pi \zeta n^2)=\sum _{n=-\infty}^{\infty} g_n\psi _p (\xi ,
%\zeta)~.
%\end{equation}
%........................
Furthermore, by discretizing the optical propagation axis $\zeta$ by
means of rational numbers, one can write the rational paraxial field
as a shifted delta comb affected by phase factors, which is the main
result of Berry and Klein:
%.......................
\begin{equation}\label{rational}%.................................Equation 5
\psi _{\rm p}\left(\xi , \frac{p}{q}\right)=\frac{1}{q^{1/2}}\sum
_{n=-\infty}^{\infty}\delta\left(\xi _{p}-\frac{n}{q}\right)\exp
[i\Phi _{\rm diffr}(n;q,p)]~,
\end{equation}
where $\xi _p=\xi -e_p/2$ and $e_p = 0$(1) if $p$ is even (odd). The
paraxial phases $\exp[i\Phi_{\rm diffr}(n;q,p)]$ are specified in
Section 4 and appear to be the physical quantities directly
connected to number theory.
At the same time, %Eq.~(\ref{rational})
this rational approximation allows for the following physical
interpretation of the paraxial self-imaging process: {\em in each
unit cell of the plane $p/q$, $q$ images of the grating slits are
reproduced with spacing $a/q$ and intensity reduced by $1/q$}.

%333333333333333333333333333333333333333333333333333333333333333333333333333333333333333333333333333333333333333333333333333333333333333333

%,,,,,,,,,,,,,,,,,,,,,,,,,,,,,,,,,,,,,,,,,,,,,,,,,,,,,,,,,.....SECTION 3                                        WAVE PROPAGATION IN OPTICAL FIBERS
\section{Wave propagation in optical fibers}
%................................................
Field dispersion phenomena in dielectric media are described in
terms of wave equations with sources of the form
%.....................................................
\begin{equation}\label{eq:prop1}
\nabla^2{\bf E}-\frac{1}{c^2}\frac{\partial^2 {\bf E}}{\partial
t^2}=\mu_0\frac{\partial^2{\bf P}}{\partial t^2}
\end{equation}
%.........................................................
that can be obtained from the Maxwell equations under minimal
assumptions on the constitutive equations. As known, in fiber optics
technology, electromagnetic dispersion is defined in terms of the
{\em propagation constant} (wavenumber) of each frequency mode
$\beta(\omega) =n(\omega) \frac{\omega}{c}$. In the following we
will use one of the simplest optical fibers having a core-cladding
step profile of the index of refraction. In addition, the famous
slowly varying envelope approximation (henceforth SVEA) is a
realistic approach when treating the propagation of
quasi-monochromatic fields, such as laser fields and other types of
coherent beams within such materials. For more details we refer the
reader to textbooks.\cite{agr}
\medskip

%\subsection{SVEA}

{\em SVEA} means decomposing the electromagnetic fields in two
factors: a rapidly varying phase component and a slowly varying
amplitude field $A$ enveloping the rapid oscillatory fields. The
following Schr\"odinger-like dispersion equation can be obtained for
$A$ in the {\em SVEA} approximation
%..........................
\begin{equation}\label{eq:prop4}
2i\frac{\partial A}{\partial z}=-{\rm sign}(\beta_2)\frac{\partial^2
A}{\partial {\tilde t}^2}~,
\end{equation}
%...........................
where $\beta _2$ is the second coefficient in the Taylor expansion
of the propagation constant in the neighbourhood of the central
resonance frequency. This is the simplest form of the dispersion
equation that one can envision in which actually no material
propagation shows up. It can be fulfilled in the practical situation
when the dielectric medium has sharp resonances ($\delta \omega
_r\ll \omega _r$). Because of technological requirements, $\beta _2$
is usually  a negative parameter corresponding to the so-called
anomalous dispersion region. As can be seen, the SVEA equation has
exactly the same mathematical form as the diffraction equation in
the paraxial approximation:
%........................................
\begin{equation}\label{eq:fresprop}
2i\frac{\partial \Psi _{\rm p}}{\partial z}=\frac{\partial^2 \Psi
_{\rm p}}{\partial x^2}~,
\end{equation}
%.......................................
where $\Psi _{\rm p}$ is the electric field close to the propagation
axis.

%44444444444444444444444444444444444444444444444444444444444444444444444444444444444444444444444444444444444444444444444444444444444444444444444444444

%,,,,,,,,,,,,,,,,,,,,,,,,,,,,,,,,,,,,,,,,,,,,,,,,,,,,,,,,,,,,,,,,SECTION 4                                          FROM DIFFRACTION TO DISPERSION
\section{From diffraction to fiber dispersion}

Many results in diffraction can be translated to the case of
dispersion in fibers by using the following substitutions
%.............................................
\[
\begin{array}{rcl}
x&\rightarrow&\tilde t\\
y&\rightarrow&{\bf r} \\
z&\rightarrow&z.
\end{array}
\]
%..............................................
In the first row one passes from the grating axis to the time axis
of a frame traveling at the group velocity of a pulse. In the second
row one passes from the second grating axis that here we consider
constant to the transverse section of the optical fiber. Finally,
the propagation axis remains the same for the two settings. This
{\it change of variables} will be used here to compare the results
obtained in the two frameworks.
%It can be readily seen that the form of both equations are similar to the Schr\"odinger equation in quantum mechanics
%...................................
%\begin{equation}\label{eq:sch}
%i\hbar\frac{\partial \Psi}{\partial t}=\frac{\hbar^2}{2m}\frac{\partial^2\Psi}{\partial x^2}+V\Psi
%\end{equation}
%................................

The general solution of the {\em SVEA} dispersion equation
~(\ref{eq:prop4}) for the amplitude $A(z,\tilde t)$ depends on the
initial conditions. Assuming a periodic input signal written as a
Fourier series, i.e., $A(0,\tilde t)=\sum
_{n=-\infty}^{n=\infty}C_n^0e^{-i\omega _n \tilde t}$, where $C_n^0$
are the Fourier coefficients of the initial pulse at the entrance of
an optical fiber with linear response, one can write the pulse at an
arbitrary $z$ as follows:
%......................................................Equation 9
\begin{equation}\label{disp_ampl_solgral}
A(z,\tilde t)=\sum C^0_n\exp{\left[i\frac{\omega^2_nz}{2}-i\omega_n
\tilde t\right]} \qquad \mbox{where  }\omega_n=2\pi n/T.
\end{equation}
%....................................................

If the scaling of variables $\tau=\tilde t/T$, $\zeta=2z/z_{{\rm
T}}$
%....................................................................
is employed, $A(z,\tilde{t})$  can be rewritten as
%...................................................
\begin{equation}\label{eq:a}
A(\zeta,\tau)=\sum C^0_n\exp{\left[i\pi n^2\zeta-i2\pi
n\tau\right]},
\end{equation}
%...................................................
because the Talbot distance corresponding to this case is $z_{\rm
T}=T^2/\pi$. Just as in the context of diffraction, the convolution
of the unitary cell with the propagator can be equally done before
or after the paraxial approximation is employed. We notice that
Eq.~\ref{eq:a} can be also written as
%.................................................Equation 11
\begin{equation}\label{dispersed_amplitude_integral}
A(\zeta,\tau)=\int^{T/2}_{-T/2}A(0, \tau')\alpha (\zeta,\tau'-\tau)d
\tau '
\end{equation}
%...............................................
since $C^0_n$ are nothing but the Fourier coefficients of the input
signal and where
%.................................................Equation 12
\begin{equation}\label{Gsum_A}
\alpha (\zeta,\tau)=\sum^{\infty}_{n=-\infty} \exp{\left[i\pi
n^2\zeta-i2\pi n\tau\right]}
\end{equation}
%................................................
can be thought of as the analog of the paraxial
propagator.\cite{ref1} In this expression, the trick is to turn the
continuous propagation axis into the rational number axis and also
to perform the integer modulo division of $n$ with respect to the
rational denominator of the propagation axis, i.e.,
%................................................Equation 13
\begin{equation}\label{rational}
\zeta=\frac{p}{q},\qquad\quad n=lq+s.
\end{equation}
%...............................
Through this approximation, the sum over $n$ is divided into two
sums: one over negative and positive integers $l$, and the other one
over $s\equiv n\,\mbox{(mod}\,q)$
%..................................................Equation 14
\begin{equation}\label{double_sum}
\alpha (\zeta,\tau)=\sum^{\infty}_{l=-\infty}\sum^{q-1}_{s=0}
\exp{\left[i\pi (lq+s)^2\frac{p}{q}-i2\pi (lq+s)\tau\right]}.
\end{equation}
%.......................................
This form of  $\alpha(\zeta,\tau)$ is almost exactly the same as
given by Berry \& Klein\cite{ref1} and by Matsutani and
\^Onishi.\cite{ref2} The difference is that the sign of the exponent
is opposite. Following these authors one can express $\alpha$ in
terms of the Poisson formula leading to
%..................................................
\begin{equation}
\alpha (\zeta,\tau)=
                  %&=&\sum^{\infty}_{l=-\infty}\sum^{q-1}_{s=0} \exp{\left[i\pi (lq+s)^2\frac{p}{q}-i2\pi (lq+s)\tau\right]} \nonumber\\
                  %&=&\sum^{q-1}_{s=0} \exp{\left[i\pi\left(\frac{p}{q}s^2-2s\tau\right)\right]\sum^{\infty}_{l=-\infty}}
                  %\exp{\left[i\pi (l^2qp+2lsp-2lq\tau)\right]} \nonumber\\
          %=&\sum^{q-1}_{s=0} \exp{\left[i\pi\left(\frac{p}{q}s^2-2s\tau\right)\right]\sum^{\infty}_{l=-\infty}}
          %\exp{\left[i\pi (lqp+2lsp-2lq\tau)\right]}\\
                               \frac{1}{\sqrt{q}}\sum _{n=-\infty}^{\infty} \Bigg[\frac{1}{\sqrt{q}}\sum _{s=0}^{q-1}
                              \exp{\left[i\pi\left(\frac{p}{q}s^2-2s\tau\right)\right]}\Bigg]\delta(\tau _p+\frac{n}{q})~,
\end{equation}
%..........................................
where $\tau _p$ is a notation similar to $\xi _p$. %$e_p =
%0$(1)if$p$is even (odd).
The rest of the calculations are straightforwardly performed though
they are lengthy. By algebraic manipulations the phase factor can be
easily obtained and we reproduce below the two expressions for
direct comparison
%.................................................Equation 15
\begin{equation} \label{phase1}
{\Phi}_{\rm disp}(n;q,p)=\frac{1}{\sqrt{q}}\sum_{s=0}^{q-1}
\exp{\left\{   i\frac{\pi}{q} \left[ps^2-2s(n-\frac{qe_p}{2})
\right] \right\}}
\end{equation}
%................................................Equation 16
\begin{equation} \label{Phase2}
 \Phi _{\rm diffr}(n;q,p)=\frac{1}{\sqrt{q}}\sum_{s=0}^{q-1} \exp{\left\{   i\frac{\pi}{q} \left[ 2s(n+\frac{qe_p}{2})-ps^2  \right] \right\}}~.
\end{equation}
%............................................
Both phases are special types of Gauss sums from the mathematical
standpoint (see the Appendix). The difference of signs here appears
because of the sign convention chosen for the Fourier transform. Not
surprisingly, the changes in the mathematical formulation are
minimal although the experimental setup is quite different.

%\bigskip

If one tries to make computer simulations using the Fourier
transform method, the Gibbs phenomenon is unavoidable for
discontinuous transmittance functions. However, in the case of fiber
dispersion, one class of continuous pulses one could work with are
the supergaussian ones, i.e., functions of the following form
%................................................Equation 17
\begin{equation}\label{superG}
A(\zeta =0,\tau) = A_0\exp \bigg[\frac{-\tau ^N}{\sigma _0}\bigg] ~,  %\qquad N\geq 2~.
\end{equation}
%...........................
where $N$ is any even number bigger than two. The bigger the chosen
$N$ the more the supergaussian pulse resembles a square pulse. A
computer simulation of the evolution of a supergaussian pulse train
is given in Fig.~2.\cite{Trev1}

%555555555555555555555555555555555555555555555555555555555555555555555555555555555555555555555555555555555555555555555555555555555555555555555

%.........................................................................................................   SECTION 5  ------  FRACTAL TALBOT
\section{Irrational Talbot distances}

\subsection{Fractal approach}

In the Talbot terminology the self-reconstructed images in the
planes $z=(p/q)z_{\rm T}$ consist of $q$ superposed copies of the
grating as already mentioned, completed with discontinuities.
Although there is a finite number of images at fractional distances,
they still represent an infinitesimal subset of all possible images
that occur at the irrational distances.

However, in the planes located at irrational fractions of the Talbot
distance the light intensity is a {\em fractal function} of the
transverse variable. The field intensity has a definite value at
every point, but its derivative has no definite value. Such fractal
functions are described by a fractal dimension, $D$, between one
(corresponding to a smooth curve) and two (corresponding to a curve
so irregular that it occupies a finite area). In the case of  Ronchi
gratings, for example, the fractal dimension of the diffracted
intensity in the irrrational transverse
planes is 3/2.\cite{ref7}  %(fig. 5).

Moreover, one can define the so-called {\em carpets} which  are
%functions of the longitudinal coordinate $z$ (along the propagation direction) as well as the transverse coordinate $x$, so the
wave intensity patterns forming surfaces on the $(\xi \, , \zeta)$
plane, i.e., the plane of the transverse periodic coordinate and the
propagation
coordinate. % (see fig. 3$b$ in Ref.~4, for example).
Since they are surfaces, their fractal dimension takes values
between two and three. According to Berry, in general for a surface
where all directions are equivalent, the fractal dimension of the
surface is one unit greater than the dimension of an inclined curve
that cuts through it. Taking into account that for a Talbot carpet
the transverse image curves have fractal dimension $D=3/2$, then the
carpet's dimension is expected to be 5/2. However, Talbot landscapes
were found not to be isotropic. For fixed $\xi$, as the intensity
varies as a function of distance $\zeta$ from
the grating, the fractal dimension is found to be 7/4, one quarter more than in the transverse case. %(fig. 5).
Therefore the longitudinal fractals are slightly more irregular than
the transverse ones. In addition, the intensity is more regular
along the bisectrix canal because of the cancelation of large Fourier components that has fractal dimension of 5/4. %(fig. 5).
The landscape is dominated by the largest of these fractal
dimensions (the longitudinal one), and so is a surface of fractal
dimension 1+ 7/4 = 11/4.

%This fractal structure also applies to the evolving wave from a particle in a box with uniform initial wave amplitude.
%There are space fractals  (varying $x$ with constant $t$), where the graph of the probability density %has $D$ = 3/2; time fractals varying $t$
%with constant $x$), where the graph of the probability density has $D$ = 7/4; and space-time fractals (along the diagonal canals), where $D$ = 5/4.
%Some of these results about %fractals can be generalized to waves evolving in enclosures of any dimensions, even with shapes that do not
%give rise to revivals, provided the initial state is discontinuous, even if only at the boundary.

%5.2  5.2  5.2  5.2

\subsection{Wavelet approach}   %............................................. SUBSECTION 5.2

Wavelet transforms (WT) are known to have various advantages over
the Fourier transform and in particular they can add up
supplementary information on the fractal features.\cite{arneodo88} A
simple definition of the one-dimensional wavelet transform is
%...........
\begin{equation}\label{wvl 0a}
W({\rm m,n})=\int _{-\infty}^{\infty} f({\rm s}) h^{*}_{{\rm
m,n}}({\rm s})d{\rm s}
\end{equation}
%...............
where $h_{{\rm m,n}}$ is a so-called daughter wavelet that is
derived from the mother wavelet $h({\rm s})$ by dilation and shift
operations quantified in terms of the dilation (m) and shift (n)
parameters:
%.............
\begin{equation}\label{wvl 0b}
h_{{\rm m,n}}({\rm s})=\frac{1}{\sqrt{{\rm m}}}h\left(\frac{{\rm
s}-{\rm n}}{{\rm m}}\right)
\end{equation}
%................
In the following, we shall use the Morlet wavelets which are derived
from the typical Gaussian-enveloped mother wavelet which is itself a
windowed Fourier transform
%.............
\begin{equation}\label{wvl 0c}
h({\rm s})=\exp[-({\rm s}/{\rm s}_0)^2]\exp (i2\pi k{\rm s}).
\end{equation}
%..............
The point is that if the mother wavelet contains a harmonic
structure, e.g., in the Morlet case the phase $\exp(i2\pi k{\rm
s})$), the WT represents both frequency and spatial information of
the signal.

In the wavelet framework one can write the expansion of an arbitrary
signal $\varphi(t)$ in an orthonormal wavelet basis in the form
%...............................................................Equation 18
\begin{equation}\label{wvl_1}
\varphi(t)=\sum _{\rm m}\sum _{\rm n} \varphi _{\rm n}^{\rm m} W
_{{\rm m,n}}(t)~,
\end{equation}
%............................
i.e., as an expansion in the dilation and translation indices, and
the coefficients of the expansion are given by
%................................................................Equation 19
\begin{equation}\label{wvl_2}
\varphi _{\rm n}^{\rm m}=\int _{-\infty}^{\infty} \varphi(t)W _{{\rm
m,n}}(t)dt~.
\end{equation}
The orthonormal wavelet basis functions $W _{{\rm m,n}}(t)$ fulfill
the following dilation-translation property
%................................................................Equation 20
\begin{equation}\label{wvl_3}
W _{{\rm m,n}}(t)=2^{{\rm m}/2}W (2^{\rm m}t-{\rm n})~.
\end{equation}
%....................

In the wavelet approach the fractal character of a certain signal
can be inferred from the behavior of its power spectrum $P(f)$,
which is the Fourier transform of the autocovariance (also termed
autocorrelation) function and in differential form $P(f)df$
represents the contribution to the variance of a signal from
frequencies between $f$ and $f+df$.
% To this
%, one can calculate spectral quantities of interest, such as the
%power spectrum, by means of wavelet transforms in order to study the
%Talbot fractal effect.
Indeed, it is known that for self-similar random processes the
spectral behavior of the power spectrum is given by\cite{wor,stas}
%........................................................................Equation 21
\begin{equation}\label{wvl_4}
P_{\varphi}(\omega)\sim |\omega|^{-\gamma _f}~,
\end{equation}
%..................
where $\gamma _f$ is the spectral parameter of the wave signal. %(not
%to be confused with the propagation constant).
In addition, the variance of the wavelet coefficients possesses the
following behavior\cite{stas}
%..................
\begin{equation}
{\rm var} \,\varphi _{\rm n}^{\rm m} \approx \left(2^{{\rm
m}}\right)^{-\gamma _f}~.
\end{equation}
%.........................

These formulas are certainly suitable for the Talbot transverse
fractals because of the interpretation in terms of the regular
superposition of identical and equally spaced grating images. We
have used these wavelet formulas in our calculations related to the
same rational paraxiallity for the two cases of transverse
diffraction fields (Fig.~3) and the fiber-dispersed optical fields
(Fig.~4), respectively. The basic idea is that the above-mentioned
formulas can be employed as a checking test of the self-similarity
structure of the optical fields. The requirement is to have a
constant spectral parameter $\gamma _f$ over many scales. In the
case of supergaussian pulses, their dispersed fields turned out not
to have the self-similarity property as can be seen by examining
Fig.~4 where one can see that the constant slope is not maintained
over all scales. In Figs.~5 and 6 the behavior of the wavelet
transform using Morlet wavelets for the diffraction field is
displayed. A great deal of details can be seen in all basic
quantities of the diffracted field, namely in the intensity,
modulus, and phase. On the other hand, the same wavelet transform
applied to the N=12 supergaussian dispersed pulse (see Fig.~7),
although showing a certain similarity to the previous discontinuous
case, contains less structure and thus looks more regular. This
points to the fact that if in diffraction experiments one uses
continuous transmittance gratings the fractal behavior would turn
milder.
%The graphics of the results are displayed in Figs.~4-6.

More realistically, paraxial waves display electric and magnetic
polarization singularities.\cite{BJOA04} If the paraxial wavefield
is treated as a signal then it is worth pointing out here that
detection of signal singularities has been studied in quite detail
by the experts in wavelet processing.\cite{Mallat92} We plan to
study this aspect in future research.

%Conclusion    Conclusion     Conclusion    Conclusion     Conclusion   Conclusion    Conclusion

%%%%.......................................................................................
\section{Conclusion}

The fractal aspects of the paraxial wavefield have been probed here
by means of the wavelet transform for the cases of diffraction and
fiber dispersion. In the case of diffraction, the previous results
of Berry and Klein are confirmed showing that the wavelet approach
can be an equivalent and more informative tool. The same procedure
applied to the case of fiber dispersion affecting the paraxial
evolution of supergaussian pulses indicates that the self-similar
fractal character does not show up in the latter type of axial
propagation. This is a consequence of the continuous transmittance
function of the supergaussian pulses as opposed to the singular one
in the case of Ronchi gratings.

Finally, as a promising perspective, we would like to suggest the
following experiment by which irrational distances can be
determined. The idea is that the spectral index of the Talbot
fractal images can be used as a very precise pointer of rational and
irrational distances with respect to the Talbot one. Suppose that
behind a Ronchi grating under plane wave illumination a CCD camera
is mounted axially by means of a precision screw. The Talbot image
at $z_{\rm T}$ can be focused experimentally and can be used to
calibrate the whole system. An implemented real time wavelet
computer software can perform a rapid determination of the fractal
index $\gamma _f$, which in turn allows the detection of changes of
the distance in order to determine if the CCD camera is at rational
or irrational multiples of the Talbot distance. To the best of our
knowledge, we are not aware of another experimental setup in which
irrational distances can be determined in such an accurate way. This
also points to high-precision applications in metrology.

%%%%................................................................................................................................
\section{Acknowledgements}

The first author wishes to thank Dr. Michel Planat for encouraging
him to study the Talbot effect from the perspective of this paper
and for discussions related to the topic. The second author would
like to thank Dr. V. Vyshloukh for the time he spent to introduce
him in the research of the self image phenomena. The referee of this
paper is gratefully acknowledged for his detailed remarks that led
to substantial improvements.

%\newpage

\bigskip
\bigskip

\noindent {\bf Appendix: Gauss sums}

\medskip

It is practically trivial to show that
%...............
\[
\sum _{n=0}^{b-1}e^{i2\pi\frac{n}{b}}=0,
\]
for any integer $b$ bigger than one.
%...............
However, similar sums of nonlinear powers in $n$ are usually nonzero
%................
\[
\sum _{n=0}^{b-1}e^{i2\pi\frac{n^k}{b}}\neq 0,
\]
%...............
where $k\geq 2$, $b$ is a prime with $b\equiv 1 \,({\rm mod} \, k)$
and the sum is over an arbitrary complete system of residues $({\rm
mod} \, b)$. Such sums are really difficult to calculate. For $k=2$,
they are known as (quadratic) Gauss sums because two centuries ago
Gauss first gave explicit results for them but only after 4 years
``with all efforts". A paper by Berndt and Evans\cite{be81} is an
excellent survey of the history of Gauss sums, see also the appendix
of the work of Merkel {\em et al} in this volume. Gauss sums having
linear phases added to the quadratic ones are of special interest in
physical applications such as the optical ones discussed in this
paper. We concentrate here on this type of Gauss sums following the
papers of Hannay and Berry\cite{hb80} and also of Matsutani and
\^Onishi.\cite{ref2}

At rational values of the propagation coordinate the diffraction and
dispersion phases can be expressed through the infinite Gaussian sum

\[G_{\infty}(a,b,c)=\lim_{N\to \infty}\frac{1}{2abN}\sum^{Nab}_{m=-Nab}\exp{\left[i\frac{\pi}{b}\left(am^2+cm\right)\right]},\]
which is an average over a period $2abN$. If $a$ and $b$ are
co-prime numbers, %i.e. $(a,b)=1$,
the sum is zero unless $c$ is an integer. Let $m=bn+s$, then the sum
is divided into two sums

\[G_{\infty}(a,b,c)=\lim_{N\to \infty}\frac{1}{2abN}\sum_{n=-Na}^{Na}\sum_{s=0}^{b-1}\exp{\left[i\frac{\pi}{b}\left(a(bn+s)^2+c(bn+s)\right)\right]},\]

%-=-=-=-=-=-=-=-=-=-=-=-=-=-=-=-=-=-=-=-=-=-=-=-=-=-=-=-=-=-=-=-=-=-=-=-=
%\foilhead[-.7in]{-2-}
%-=-=-=-=-=-=-=-=-=-=-=-=-=-=-=-=-=-=-=-=-=-=-=-=-=-=-=-=-=-=-=-=-=-=-=-=
% Set the page color per slides
%\pagecolor{red}
%\includegraphics{cover.jpg}

The argument of the exponential in the sum becomes
\[i\frac{\pi}{b}\left(a(bn+s)^2+c(bn+s)\right)=i2\pi ans+i \pi n (abn+c)+i\frac{\pi}{b}\left(as^2+bs\right)\]
 and if $ab$ and $c$ are both odd or even at the same time, only terms in $s$ survive: if $n$ is even, the term is even and if $n$ is
 odd, i.e. $n=2k_1+1$, and also $ab=2k_2+1$ and $c=2k_3+1$, the term is
%...............
\begin{eqnarray*}
i \pi n (abn+c)&=&i\pi (2k_1+1)[(2k_2+1)(2k_1+1)+(2k_3+1)]\\
                      &=&i\pi(2k_1+1)(4k_1k_2+2k_2+2k_1+2k_3+2)\\
                      &=&i2\pi(2k_1+1)(2k_1k_2+k_2+k_1+k_3+1) \\
\end{eqnarray*}

%-=-=-=-=-=-=-=-=-=-=-=-=-=-=-=-=-=-=-=-=-=-=-=-=-=-=-=-=-=-=-=-=-=-=-=-=
%\foilhead[-.7in]{-3-}
%-=-=-=-=-=-=-=-=-=-=-=-=-=-=-=-=-=-=-=-=-=-=-=-=-=-=-=-=-=-=-=-=-=-=-=-=
Thus $G_{\infty}$ is rewriten as

\[G_{\infty}(a,b,c)=\lim_{N\to \infty}\frac{1}{2abN}\sum_{n=-Na}^{Na}\sum_{s=0}^{b-1}\exp{\left[i\frac{\pi}{b}\left(as^2+cs\right)\right]},\]
and taking the limit, $G_{\infty}$ coincides with the quadratic
Gauss sum

\[G_{\infty}(a,b,c)=S(a,b,c)=\frac{1}{a}\sum^{a-1}_{m=0}\exp{\left[i\frac{\pi}{b}\left(am^2+cm\right)\right]}.\]
Notice the following property: in the normalization fraction and the
upper limit of the sum, $a$ can be interchanged for $b$
simultaneously. We will use this property because it is appropriate
from the standpoint of the physical application
to run the sum over the denominator of the phase, i.e., over $b$. %The implications will be analyzed in detail.
%-=-=-=-=-=-=-=-=-=-=-=-=-=-=-=-=-=-=-=-=-=-=-=-=-=-=-=-=-=-=-=-=-=-=-=-=
%\foilhead[-.7in]

\medskip

%\medskip
We consider now the various cases according to the odd even
character of the parameters $a$, $b$ and $c$. Only the first case
will be treated in detail.

\medskip

%-=-=-=-=-=-=-=-=-=-=-=-=-=-=-=-=-=-=-=-=-=-=-=-=-=-=-=-=-=-=-=-=-=-=-=-=
\underline{Case $a$ and $c$ even, $b$ odd (because of coprimality to
$a$)}

\medskip

\noindent A factor $\exp{(i\pi cmk)}$ can be added since $c$ is
even. Then
%........................
\[
S(a,b,c)%&=&\frac{1}{a}\sum^{a-1}_{m=0}e^{i\frac{\pi}{b}\left(am^2+cm\right)}\\
        =\frac{1}{b}\sum^{b-1}_{m=0} e^{i\frac{\pi}{b}am^2}
e^{i\frac{\pi}{b}cm(kb+1)}
%\frac{1}{a}\sum^{a-1}_{m=0}e^{i\frac{\pi}{b}am^2}e^{i\frac{\pi}{b}cm}.\\
\]
%.......................
%Since $c$ is even, a factor $\exp{(i\pi cmk)}$ can be added.

%\[
%S(a,b,c)=\frac{1}{a}\sum^{a-1}_{m=0} e^{i\frac{\pi}{b}am^2}
%e^{i\frac{\pi}{b}cm(kb+1)}
%\]

%-=-=-=-=-=-=-=-=-=-=-=-=-=-=-=-=-=-=-=-=-=-=-=-=-=-=-=-=-=-=-=-=-=-=-=-=
%\foilhead[-.7in]{}
%-=-=-=-=-=-=-=-=-=-=-=-=-=-=-=-=-=-=-=-=-=-=-=-=-=-=-=-=-=-=-=-=-=-=-=-=
Noticing that $kb+1\equiv 1({\rm mod}\,b)$ and defining $a\bar
a_b\equiv 1({\rm mod}\,b)$ we can write
\begin{eqnarray*}
S(a,b,c) %&=&\frac{1}{a}\sum^{a-1}_{m=0}e^{i\frac{\pi}{b}am^2}e^{i\frac{\pi}{b}acm\bar a_b}\\
        =\frac{1}{b}\sum^{b-1}_{m=0}\exp\left(\left[i\pi\frac{a}{b}\left(m^2+cm\bar
        a_b\right)\right]\right)
\end{eqnarray*}
and completing the square in the argument of the exponential:
\begin{eqnarray*}
S(a,b,c)%&=&\frac{1}{b}\sum^{b-1}_{m=0}\exp{\left[i\pi\frac{a}{b}\left(m^2+cm\bar a_b\right)\right]}\\
        &=&\frac{1}{b}\sum^{b-1}_{m=0}\exp\left(i\pi\frac{a}{b}\left[\left(m+\frac{c\bar a_b}{2}\right)^2-
        \left(\frac{c\bar a_b}{2}\right)^2\right]\right)
\end{eqnarray*}

Next, one should notice the following simpler form:
\[
\sum_{m=0}^{b-1} \exp{ \left[ i\pi \frac{a}{b}\left(m+\frac{c\bar
a_b}{2}\right)^2  \right]
}=\sum^{b-1}_{n=0}\exp{\left(i\pi\frac{a}{b}n^2\right)}
\]
since the linear term plus the independent one in the left hand side
exponential lead to a different order of the terms with respect to
the right hand side of the equation. Furthermore we note that

\[
\sum_{n=0}^{b-1}\exp{\left(i\pi\frac{a}{b}n^2\right)}=\sum_{n=0}^{b-1}
\left[1+\left({an/2}\over{b}\right)_L\right]\exp{\left(i\pi\frac{a}{b}n\right)}
\]
where $\left({l_1}\over{l_2}\right)_L$ denotes the Legendre symbol
that for $l_2$ an odd prime number is defined as zero if $l_1$ is
divided by $l_2$, 1 if $l_1$ is a quadratic residue mod $l_2$ (there
exists an integer $k$ such that $k^2=l_1$ (mod $l_2$)), and -1 if
$l_1$ is not a quadratic residue.

%-=-=-=-=-=-=-=-=-=-=-=-=-=-=-=-=-=-=-=-=-=-=-=-=-=-=-=-=-=-=-=-=-=-=-=-=
%\foilhead[-.7in]{.}
%-=-=-=-=-=-=-=-=-=-=-=-=-=-=-=-=-=-=-=-=-=-=-=-=-=-=-=-=-=-=-=-=-=-=-=-=
Therefore the Gauss sum is expressed as follows:
\begin{eqnarray*}
S(a,b,c)
%&=&\frac{1}{b}\sum_{n=0}^{b-1}\exp{\left\{i\pi\frac{a}{b}\left[\left(m+\frac{c\hat a_b}{2}\right)^2-\left(\frac{c\hat a_b}{2}\right)^2\right]\right\}}\\
 &=&\frac{1}{b}\exp{\left(-i\pi\frac{a}{b}(c\bar a_b/2)^2\right)}
 \sum_{n=0}^{b-1} \left[1+\left(\frac{an/2}{b}\right)_L\right]\exp{\left(i\pi\frac{a}{b}n\right)}\\
\end{eqnarray*}
and the latter sum under splitting into two sums yields zero in the
first one. Making use of the Legendre symbol multiplicative property
in the top argument we get:

\[
S(a,b,c)=\frac{1}{b}\left(\frac{a/2}{b}\right)_L\exp{\left(-i\pi\frac{a}{b}(c\bar
a_b/2)^2\right)}\sum_{n=0}^{b-1}\left(\frac{n}{b}\right)_L\exp{\left(i2\pi\frac{n}{b}\right)}.
\]

The following result from number theory\cite{koblitz} should be used
in order to get the most compact result for $S(a,b,c)$:

\[\left(\frac{1/2}{b}\right)_L=\left(\frac{2}{b}\right)_L=(-1)^{(b^2-1)/8}=\exp{\left(\pm i\pi\frac{b^2-1}{8}\right)}\]

and then
\[
\sum_{n=0}^{b-1}\left(\frac{n}{b}\right)_L\exp{\left(i2\pi\frac{n}{b}\right)}=\sqrt{b}\exp{\left(
i\pi\frac{(b-1)^2}{8}\right)}\]

for $b\equiv 1\,({\rm mod}\, 4)$ or $b\equiv 3\,({\rm mod}\, 4)$.
Then:
%.......................
\begin{eqnarray*}
S(a,b,c)&=&\frac{1}{b}\left(\frac{a}{b}\right)_L\left(\frac{1/2}{b}\right)_L\exp{\left(-i\pi\frac{a}{b}(c\bar
a_b/2)^2\right)}
\sum_{n=0}^{b-1}\left(\frac{n}{b}\right)_L\exp{\left(i2\pi\frac{n}{b}\right)}\\
    &=&\frac{1}{\sqrt{b}}\left(\frac{a}{b}\right)_L\exp{\left(\pm i\pi\frac{\left(b^2-1\right)}{8}-i\pi\frac{a}{b}(c\bar a_b/2)^2 +
    i\pi\frac{(b-1)^2}{8}\right)}.\\
\end{eqnarray*}

The terms in the exponential with the common factor $i\pi/8$ are
rewritten

\[\frac{i\pi}{8} \left(\pm b^2\mp1+b^2-2b+1\right)=-\frac{i2\pi}{8}(b-1)\]
%...........................................
having chosen the lower sign (there is no final change if the $+$
sign is picked up).

%-=-=-=-=-=-=-=-=-=-=-=-=-=-=-=-=-=-=-=-=-=-=-=-=-=-=-=-=-=-=-=-=-=-=-=-=
%\foilhead[-.7in]{}
%-=-=-=-=-=-=-=-=-=-=-=-=-=-=-=-=-=-=-=-=-=-=-=-=-=-=-=-=-=-=-=-=-=-=-=-=
Recalling now that $a$ and $c$ are even we get

\[S(a,b,c)=\frac{1}{\sqrt{b}}\left(\frac{a}{b}\right)_L\exp\left(- i\pi\left[\frac{1}{4}\left(b-1\right)+\frac{a}{b}(c\bar a_b/2)^2\right]\right).\]

{\underline{Case $a$, $b$, and $c$ odd}

\medskip

\noindent Exactly the same expression can be worked out with the
only minor modification that $a$ is \textit{replaced} by $2
\bar{2}_ba$.

%\[S(a,b,c)=\frac{1}{\sqrt{b}}\left(\frac{a}{b}\right)_L
%\exp\left(- i\pi\left[\frac{1}{4}\left(b-1\right)+\frac{2a}{b}\bar
%2_b \left(c{\overline{2a}}_b\right)^2\right]\right).\]
%........................................
%and the cases $a$ even, $b$ odd; and $a$ odd, $b$ odd.

\underline{Case $a$ odd, $b$ even, $c$ odd}

\medskip

\noindent Similar arguments lead to a slightly different result.

%-=-=-=-=-=-=-=-=-=-=-=-=-=-=-=-=-=-=-=-=-=-=-=-=-=-=-=-=-=-=-=-=-=-=-=-=
%\foilhead[-.7in]{}
%-=-=-=-=-=-=-=-=-=-=-=-=-=-=-=-=-=-=-=-=-=-=-=-=-=-=-=-=-=-=-=-=-=-=-=-=

%\newpage

\noindent The three final results are as follows:

\[S(a,b,c)=\frac{1}{\sqrt{b}}\left(\frac{a}{b}\right)_L\exp\left(- i\frac{\pi}{4}\left[\left(b-1\right)+\frac{a}{b}(c\bar
a_b)^2\right]\right) \qquad \quad \, \, \mbox{ $a$ even, $b$ odd,
$c$ even},\]

\[S(a,b,c)=\frac{1}{\sqrt{b}}\left(\frac{a}{b}\right)_L\exp\left(- i\frac{\pi}{4}\left[\left(b-1\right)+\frac{2a}{b}\bar 2_b^3 \left(c\bar{a}_b
\right)^2\right]\right)\quad \mbox{ $a$ odd, $b$ odd, $c$ odd},\]

\[S(a,b,c)=\frac{1}{\sqrt{b}}\left(\frac{b}{a}\right)_L\exp\left(- i\frac{\pi}{4}\left[-a+\frac{a}{b} (c\bar
a_b)^2\right]\right) \qquad \qquad \quad\, \mbox{ $a$ odd, $b$ even,
$c$ odd}.\]

These results were obtained under the assumption that $b$ is prime.
For the general case where $b$ is not a prime the Legendre symbol
should be substituted by the Jacobi symbol, which is a product of
Legendre symbols defined in the following way. Let the prime
decomposition of $b$ be $b=\prod_{i=1}^{i=n} p_{i}^{r_i}$. Then, by
definition, the Jacobi symbol is

\[
\left(\frac{a}{b}\right)_J=\prod
_{i=1}^{i=n}\left(\frac{a}{p_i}\right)^{r_i}_{L}.
\]

In this way, the evaluation of the Talbot phases are performed by
identification of the parameters: $a=\pm p$, $b=q$, $c=\mp 2(n\mp
qe_p/2)$, respectively, where $e_p=0$ (1) if $p$ even (odd). The
upper sign is for dispersion and the lower for diffraction. Thus, we
get:

\medskip

$p$ even, $q$ odd:
\begin{eqnarray*}
\Phi_{{\rm disp}}(n;p,q)&=&\left(\frac{p}{q}\right)_J\exp\left(-
i\frac{\pi}{4}\left[\left(q-1\right)+\frac{p}{q}(2n\bar
p_q)^2\right]\right) ,\\
\Phi_{{\rm diffr}}(n;p,q)&=&\left(\frac{p}{q}\right)_J\exp\left(+
i\frac{\pi}{4}\left[\left(q-1\right)+\frac{p}{q}(2n\bar
p_q)^2\right]\right).
\end{eqnarray*}

\medskip

$p$ odd, $q$ odd:
\begin{eqnarray*}
\Phi_{{\rm disp}}(n;p,q)&=&\left(\frac{p}{q}\right)_J\exp\left(-
i\frac{\pi}{4}\left[\left(q-1\right)+2\bar
2_q^3\frac{p}{q}((2n-q)\bar
p_q)^2\right]\right) ,\\
\Phi_{{\rm diffr}}(n;p,q)&=&\left(\frac{p}{q}\right)_J\exp\left(+
i\frac{\pi}{4}\left[\left(q-1\right)+2\bar
2_q^3\frac{p}{q}((2n+q)\bar p_q)^2\right]\right).
\end{eqnarray*}

\medskip

$p$ odd, $q$ even:
\begin{eqnarray*}
\Phi_{{\rm disp}}(n;p,q)&=&\left(\frac{q}{p}\right)_J\exp\left(-
i\frac{\pi}{4}\left[-p+\frac{p}{q}((2n-q)\bar
p_q)^2\right]\right) ,\\
\Phi_{{\rm diffr}}(n;p,q)&=&\left(\frac{q}{p}\right)_J\exp\left(+
i\frac{\pi}{4}\left[-p+\frac{p}{q}((2n+q)\bar p_q)^2\right]\right).
\end{eqnarray*}

\newpage
\centerline{{\bf Figure captions}}

\bigskip

\noindent {\bf FIG. 1}: Image of a Ronchi grating as obtained on a
photographic plate located at the Talbot distance $z_T=28.4$ cm.
This and many similar images have been obtained in the graduate work
of Trevi\~no-Guti\'errez\cite{Trev1} in which a He-Ne laser working
at its usual wavelength $\lambda =632.8$ nm has been used.

\medskip

\noindent {\bf FIG. 2}: Computer simulation of $|A(\zeta, \tau)|^2$
of a $N=12\, , \sigma _0=1.5$ supergaussian pulse train as given in
Eq.~(\ref{superG}) in a linear fiber characterized by the
Schr\"odinger-like dispersion relation given in
Eq.~(\ref{eq:prop4}).

\medskip

\noindent {\bf FIG. 3}: (a) The fractal Talbot light intensity
$|\Psi _{\rm p}|^2$ at $\zeta =144/377$ and (b) the plot of the
logarithmic variance of its wavelet coefficients
(Eq.~(\ref{wvl_2})). The line of negative slope of the latter
semilog plot indicates fractal behaviour of the diffraction
wavefield as we expected. The fractal coefficient is given by the
slope and its calculated value is $\gamma _f$.

\medskip

\noindent {\bf FIG. 4}: Snapshot of the dispersed supergaussian
pulse for $N=12$ at $\zeta =144/377$ (close to the Golden Mean
$\zeta _G =3/2-\sqrt{5}/2$).
The log variance plot is monotonically decreasing %without any local maximum and
%only
displaying a plateau indicating a nonfractal behaviour of the $N=12$
supergaussian pulse train.

\medskip

\noindent {\bf FIG. 5}: The wavelet transform of the intensity
$|\Psi _{{\rm p}}|^2$ at $\zeta =144/377$ for (a) the unit cell and
(b) half-period displaced grating unit cell. There is no difference
because the square modulus is plotted.
%and its fractal coefficient $\beta _F$.

\medskip

\noindent {\bf FIG. 6}: Wavelet representations of: (a) the squared
modulus of the amplitude and (b) phase of the Talbot diffraction
field for fixed $\zeta = 144/377$ and a displaced unit cell.

\medskip

%FIG. 7: The wavelet transform of (a) the amplitude $A$ at $\zeta
%=144/377$ and (b) the intensity of a N=12 supergaussian pulse in the
%unit cell.
%and its fractal coefficient $\beta _F$.

\medskip

\noindent {\bf FIG. 7}: Wavelet representations of the (a) amplitude
and (b) phase of the Talbot dispersed supergaussian field ($N=12$)
for $\zeta = 144/377$.

\end{document}